\documentclass[10pt]{article}
\usepackage{amsmath}
\usepackage{graphicx}

\title{MACROSCOPIC QUANTUM GAME}
\author{{\bf Andrey
Grib\footnote{e-mail: grib@friedman.usr.lgu.spb.su}}\\ {\bf
Georges Parfionov\footnote{e-mail: your@GP5574.spb.edu}} \\
{\small\em Alexandre Friedmann Laboratory of Theoretical
Physics,} \\ {\small\em St.Petersburg University of
Economics and Finances} \\ {\small\em St.Petersburg,Russia
191023} \\ {\small\em Fax: +7(812)110-5742,\; Telephone:
+7(812)110-5605}}
\date{}

\begin{document}
\maketitle
\begin{abstract}

\noindent The game in which acts  of participants don't
have an adequate description in terms of  Boolean logic and
classical theory of probabilities is considered. The model
of the game interaction is constructed on the basis of a
non-distributive orthocomplemented lattice. Mixed
strategies of the participants are calculated by the use of
probability amplitudes according to the rules of quantum
mechanics. A scheme of quantization of the payoff function
is proposed and an algorithm for the search of Nash
equilibrium is given. It is shown that differently from the
classical case in the quantum situation a discrete set of
equilibria is possible.
\end{abstract}

It often occurs that mathematical structures discovered
when solving some class of  problems find their natural
application in totally different areas. The mathematical
formalism of quantum mechanics operating with such notions
as "observable", "state", "probability amplitude" is not an
exception to this rule. The goal of the present paper is to
show that the language of quantum mechanics, initially
applied to the description of the microworld, is adequate
for the description of some macroscopic systems and
situations where Planck's constant plays no role. It is
natural to look for applications of the formalism of
quantum mechanics in those situations when one has
interactions with the element of indeterminacy. In
\cite{Gribook} as well as more recently \cite{Waldir} it
was shown that the quantum mechanical formalism can be
applied to description of macroscopical  systems when {\it
the distributive} property for random events is broken. In
the physics of the microworld non-distributivity has an
objective status and must be present in principle. For
macroscopic systems the non-distributivity of random events
expresses some specific case of the observer's "ignorance".

In the present paper a quantum mechanical formalism is
applied to the analysis of a  conflict interaction, the
mathematical model for which is an antagonistic game of two
persons. The game is based on a generalization of examples
of the macroscopical automata simulating  the behaviour of
some quantum systems considered earlier
in~\cite{GrRZ1,GrRZ2}. A special feature of the game
considered is that the players acts  go in contradiction
with the usual logic. The consequence is breaking of the
classical probability interpretation of the mixed strategy:
the sum of the probabilities for alternate outcomes may be
larger than one. The cause of breaking of the basic
property of the probability is in the {\it
non-distributivity of the logic}. The partners relations
are such that the disjunction "or", conjunction "and" and
the operation of negation do not form  a Boolean algebra
but an orthocomplemented non-distributive lattice. However
this ortholattice happens to be just that which  describes
some properties of a quantum system with spin one half.
This leads to new "quantum" rules for the calculations of
the average profit and new representation of the mixed
strategy, the role of which is played by the "wave
function" -- the normalized vector in a finite dimensional
Hilbert space. Calculations of probabilities are made
according to the standard rules of quantum mechanics.
Differently from the examples of quantum games considered
in~\cite{Eisert,Ekert,Marinatto} where the "quantum" nature
of the game was conditioned by the microparticles or
quantum computers based on them,in our case we deal with a
{\it macroscopic} game, the quantum nature of which has
nothing to do with microparticles. This gives the hope that
our example is one of many analogous situations in biology,
economics etc where the formalism of quantum mechanics can
be used.

The game "Wise Alice" formulated in our paper is a
modification of the well known game when each of the
participants names one of some previously considered
objects. In the case if the results differ, one of the
players wins from the other some agreed sum of money. The
participants of our game A and B, call them Alice and Bob
have a quadratic box in which a ball is located. Bob puts
his ball in one of the corners of the box but doesn't tell
his partner which corner.

\begin{figure}[ht]
\begin{center}
\begin{picture}(100,130)
\linethickness{1pt}

\put(25,115){\circle{15}}
\put(80,115){\vector(-1,0){30}}

\put(0,110){\bf 1} \put(100,110){\bf 2} \put(100,15){\bf 3}
\put(0,15){\bf 4}

\linethickness{2pt}

\put(-16,130){\line(1,0){132}} \put(-16,0){\line(1,0){132}}
\put(-15,0){\line(0,1){130}} \put(115,0){\line(0,1){130}}
\end{picture}
\end{center}
\vspace{-15pt}\caption{\em Bob's ball moves into the place
asked by Alice} \label{fig1}
\end{figure}
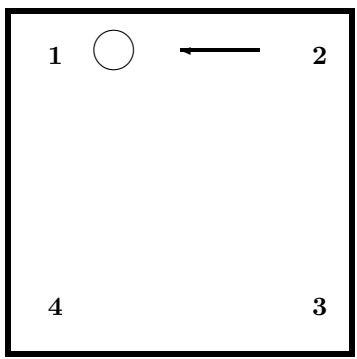

\noindent Alice must guess in which corner Bob has put his
ball. The rules of the game are such that Alice can ask Bob
questions supposing the two-valued answer: "yes" or "no".
It is supposed that Bob is honest and always tells the
truth. In the case of a "yes" answer Alice is satisfied, in
the opposite case she asks Bob to pay her some
compensation. However, differently from other such
games~\cite{Moulin,Owen} the rules of this game (see
Figure~\ref{fig1}) have one specific feature: {\it Bob has
the possibility to move the ball to any of the adjacent
vertices of the square after Alice asks her question.} This
additional condition decisively changes the behaviour of
Bob, making him to become active under the influence of
Alice's questions. Due to the fact that negative answers
are not profitable for him he, in all possible cases, moves
his ball to the convenient adjacent vertex.

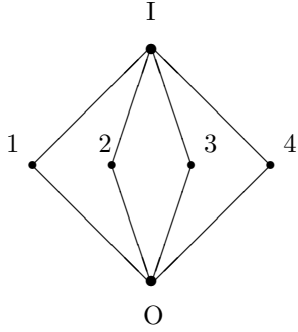
\begin{figure}[ht]
\begin{center}
\begin{picture}(120,120)
\put(45,94){\circle*{4}} \put(43,105){I}
\put(0,50){\circle*{3}} \put(-10,55){1}
\put(30,50){\circle*{3}} \put(25,55){2}
\put(60,50){\circle*{3}} \put(65,55){3}
\put(90,50){\circle*{3}} \put(95,55){4}
\put(45,6){\circle*{4}} \put(42,-10){O}
\put(0,50){\line(1,1){45}} \put(0,50){\line(1,-1){45}}
\put(90,50){\line(-1,1){45}} \put(90,50){\line(-1,-1){45}}
\put(30,50){\line(1,3){15}} \put(30,50){\line(1,-3){15}}
\put(60,50){\line(-1,3){15}} \put(60,50){\line(-1,-3){15}}
\end{picture}
\end{center}
\caption{\em Lattice of Alice's questions and
Bob's answers} \label{fig3}
\end{figure}

So being in vertices 2 or 4 and getting from Alice the
question "Are you in the vertex 1?" Bob quickly puts his
ball in the asked vertex and honestly answers"yes".
However, if the Bob's ball was initially in the vertex 3 he
cannot escape the negative answer notwithstanding to what
vertex he moves his ball and he fails. One must pay
attention that in this case Alice not only gets the profit
but also obtains the {\it exact information} on the initial
position of the ball: Bob's honest answer immediately
reveals his initial position. The interaction of our
players can be described by a four on four matrix
$~(h_{ik})~$  representing payoffs of Alice in each of the
16 possible game situations
\begin{table}[h]
\begin{center}
 \begin{tabular}{|c||c|c|c|c|}
\hline $A \backslash B$ &   {\bf 1} & {\bf 2} & {\bf 3} &
{\bf 4} \\ \hline\hline
 {\bf ~1~} & ~0~ &  ~0~ & ~a~ & ~0~ \\ \hline
 {\bf ~2~} & ~0~ &  ~0~ & ~0~ & ~b~ \\ \hline
 {\bf ~3~} & ~c~ &  ~0~ & ~0~ & ~0~ \\ \hline
 {\bf ~4~} & ~0~ &  ~d~ & ~0~ & ~0~ \\ \hline
 \end{tabular}
\end{center}\caption{\em The Payoff-matrix of Alice}\label{h1}
\end{table}
\noindent where $~a,b,c,d>0~$ are her payoffs in those
situations when Bob cannot answer her questions
affirmatively. Our game is an antagonistic game,so the
payoff matrix of Bob is the opposite to that of Alice:
$(-h_{ik})$. The main problem of  game theory is to find
so-called {\it points of equilibrium or saddle points} --
game situations, optimal for all players at once. It is
easy to see that the classical game with our payoff matrix
does not have such equilibrium points. Nonexistence of the
saddle point follows from the strict inequality valid for
our game $$\max_{j}\min_{k}h_{jk}< \min_{k}\max_{j}h_{jk}$$
So there are no stable strategies to follow for Bob and
Alice in each {\it separate} turn of the game. In spite of
the absence of a rational choice at each turn of the game,
when the game is repeated many times some optimal lines of
behaviour can be found. To find them in the theory of
classical games one must, following von
Neumann~\cite{Neumann}, look for the so called mixed
generalization of the game. The optimal mixed strategies
for Alice and Bob are defined as such probability
distributions on the sets of pure strategies
$x^0=(x^0_1,x^0_2,x^0_3,x^0_4 )$ and
$y^0=(y^0_1,y^0_2,y^0_3,y^0_4)$ that for all distributions
of $~x, y~$ the von Neumann-Nash inequalities are valid:
\begin{equation}\label{NE}
       {\cal H}_A (x^0,y^0)\geq{\cal H}_A (x,y^0)\,,\qquad {\cal H}_B
       (x^0,y^0)\geq{\cal H}_B (x^0,y),
\end{equation}
where { \it$~\cal{H}_{A},\cal{H}_{B}~$ -- payoff functions}
of Alice and Bob are the expectation values of their wins
$${\cal H}_A(x,y) = \sum_{j,k=1}^{4}{h_{jk} x_j y_k}\,,
\qquad {\cal H}_B(x,y) = -\sum_{j,k=1}^{4}{h_{jk}x_j y_k}$$
The combination of strategies, satisfying the von
Neumann-Nash inequalities, is called {\it the situation of
equilibrium} in Nash's sense. However in the case of our
game the logic of behaviour of the players is such that
usual classical theory does not work. To see this consider
Hasse diagram (Fig.~\ref{fig3}) where to atoms correspond
different possibilities for Bob from the point of view of
Alice when she pays attention only to his negative
answers.One has the special structure of disjunction so
that for example 1 or 2 is true when 1 true or 2 true but
not always true.
\noindent If one considers all outcomes equally possible,
then the probability of the always true event, i.e.
disjunction of any of two events occurs to be one half! The
distributivity property is broken. So a classical
probabilistic description of the behaviour of the players
in the repeated game is impossible in principle.

The solution for the situation arising is given by the
ideas of quantum mechanics. Following A.A.Grib and
R.R.Zapatrin~{\cite{GrRZ1} we pay attention to the fact
that the ortholattice of the logic of interaction of
partners of the "Wise Alice" is isomorphic to the
ortholattice of invariant subspaces of the Hilbert space of
the quantum system with spin $\frac{1}{2}$ and observables
of the type of $S_x$ $S_\theta$. As it is well known one
can represent this lattice by considering on the plane two
pairs of mutually orthogonal direct lines $\{a_1; a_3\}$,
$\{a_2; a_4\}$. One of these pairs makes diagonal the
operator $S_x$, the other $S_\theta$. If one takes as
representations of logical conjunction and disjunction
their intersection and linear envelope and if negation
corresponds to the orthogonal complement one obtains the
ortholattice isomorphic to the logic of our players.  We
saw that in one "experiment" neither Alice nor Bob have a
stable strategy. However if the game is repeated many times
one can ask about optimal frequencies of the corresponding
pure strategies. Due to the non-distributivity of the logic
it is impossible to define on the sets $S_A$ and $S_B$ of
pure strategies a probabilistic measure. The main problem
is calculation of an adequate {\it procedure of averaging}.
Following well known constructions of quantum mechanics we
take instead of the sets of pure strategies of Alice and
Bob $S_A, S_B$ the pair of two-dimensional Hilbert spaces
$H_A, H_B$. So {\it pure strategies} are represented by
one-dimensional subspaces or {\it normalized vectors} of
Hilbert space (wave functions). Use of Hilbert space
permits us without any difficulties to realize the
non-distributive logic of our players. So the average
payoff for the given types of behaviour of the players:
$$E_{\varphi\otimes\psi}\widehat{{\cal H}}_A
=\sum_{j,k=1}^{4}{h_{jk}
\langle\widehat{\alpha}_j\varphi,\, \varphi\rangle \cdot
\langle\widehat{\beta}_k\psi,\, \psi\rangle }$$ Putting
into this formula the elements of our payoff matrix and
using the notations
\mbox{$~p_j=\langle\widehat{\alpha}_j\varphi,\,\varphi\rangle~$},
$~q_k=\langle\widehat{\beta}_k\psi,\,\psi\rangle~$ one
obtains
\begin{equation}\label{Hquant}
E_{\varphi\otimes\psi}\widehat{{\cal H}}_A =ap_1 q_3 +cp_3
q_1 + bp_2 q_4 + dp_4 q_2
\end{equation}
The definition of the Nash equilibrium for the quantum case
is not much different from the classical case~(\ref{NE})
and can be written as $$E\widehat{{\cal H}}_A
(\varphi^0,\psi^0)\geq E\widehat{{\cal H}}_A
(\varphi,\psi^0),\qquad E\widehat{{\cal H}}_B
(\varphi^0,\psi^0)\geq E\widehat{{\cal H}}_B
(\varphi^0,\psi)$$ It is convenient to find the equilibrium
points in the coordinate form. To do this let us fix in the
space of strategies of Alice $~H_A~$ eigenbasis $\{\xi_1^+,
\xi_1^-\}$ ¨ $ \{\xi_2^+, \xi_2^-\}$ corresponding to two
projectors $\widehat{\alpha}_1, \widehat{\alpha}_2$ and let
us do the same for Bob, taking bases $\{\eta_1^+,
\eta_1^-\}$ ¨ $ \{\eta_2^+, \eta_2^-\}$. The angles between
the largest eigenvectors denote as $\theta_A$ and
$\theta_B$. Then one can write in the quantum payoff
function $$E\widehat{{\cal H}}_A({\varphi, \psi}) =ap_1 q_3
+cp_3 q_1 + bp_2 q_4 + dp_4 q_2 $$ the squares of moduli of
the amplitudes $~p_j,p_k~$ as $$ p_1=\cos^2\alpha,\quad
p_3=\sin^2\alpha,\quad p_2=\cos^2(\alpha-\theta_A),\quad
p_4=\sin^2(\alpha-\theta_A), $$
    $$q_1=\cos^2\beta,\quad q_3=\sin^2\beta,\quad
    q_2=\cos^2(\beta-\theta_B),\quad q_4=\sin^2(\beta-\theta_B),$$
where $\alpha$ ¨ $\beta$  are the angles of vectors
$\varphi, \psi$  to the corresponding axises. For values of
angles one can take the interval $[0^0; 180^0]$. In the
result the problem of search of the equilibrium points of
the quantum game became the problem of finding a minimax of
the function of two angle variables $$F(\alpha, \beta)=
a\cos^2\alpha \sin^2\beta +c\sin^2\alpha \cos^2\beta +$$
$$+b\cos^2(\alpha-\theta_A) \sin^2(\beta-\theta_B) +
d\sin^2(\alpha-\theta_A) \cos^2(\beta-\theta_B)$$ on the
square $[0^0; 180^0]\times[0^0; 180^0]$. Differently from
the geometrical saddle points the conditions of the Nash
equilibrium are not just putting to zero values of the
corresponding partial derivatives. So in the situation of
absence of simple analytical solutions one must look for
numerical methods. To do calculations we use an algorithm
based on the construction of "curves of reaction" or
"curves of the best answers" of the participants of the
game. The definition of  curves of reaction is based on the
following consideration. If Alice knew what decision Bob
will take  she could make an {\it optimal} choice. But the
essence of the game situation is that she doesn't know
it.She must take into account his different strategies and
on each possible act of the partner she must find the
optimal way to act. Her considerations look like
considerations of the player,expressed by the formula: "if
he does this, then I shall do that". Bob thinks the same
way. So one must consider two functions, $$\alpha = {\cal
R}_A (\beta)\quad \mbox{and}\quad \beta = {\cal R}_B
(\alpha)$$ the plots of which are called the curves of
reactions of Alice and Bob. Due to the definition of these
functions $$\max_{\lambda} F(\lambda,\, \beta)= F({\cal
R}_A (\beta),\, \beta), \quad \min_{\mu} F(\alpha,\,
\mu)=F(\alpha,\, {\cal R}_B (\alpha))$$ It is easy to see
that intersections of curves of reaction give points of
Nash equilibrium. Numerical experiments show that dependent
on the values of the parameters $~a,b,c,d~$of the payoff
function and the angles characterizing the type of
 player one has qualitatively different pictures.
Intersections can be absent, there can be one intersection
and lastly there can be the case with two equilibrium
points with different values of the payoff of the game,
which is absent in the case of the classical matrix game.

{\bf 1. \em Two equilibrium} points arise in the case of
the payoff matrix
  \begin{table}[h]\label{h2}
\begin{center}
\begin{tabular}{|c||c|c|c|c|}
\hline $A \backslash B$ &   {\bf 1} & {\bf 2} & {\bf 3} &
{\bf 4} \\ \hline\hline
 {\bf ~1~} & ~0~ &  ~0~ & ~3~ & ~0~ \\ \hline
 {\bf ~2~} & ~0~ &  ~0~ & ~0~ & ~3~ \\ \hline
 {\bf ~3~} & ~5~ &  ~0~ & ~0~ & ~0~ \\ \hline
 {\bf ~4~} & ~0~ &  ~1~ & ~0~ & ~0~ \\ \hline
 \end{tabular}
\end{center}
\end{table}
and an operator representation of the ortholattice
corresponding to  angles $\theta_A=10^0$, $\theta_B=70^0$.
One of the equilibrium points is  inside the square, the
other one is on it's boundary (see~Fig.~\ref{fig5}). The
curves of reaction in this case happen to be {\it
discontinuous}. For convenience the discontinuities are
shown by thin lines. The discontinuous character of the
curve of reaction of Alice made it impossible  for one more
equilibrium point to occur.
\begin{figure}[ht]
\begin{center}
\includegraphics[height=0.48\textwidth]{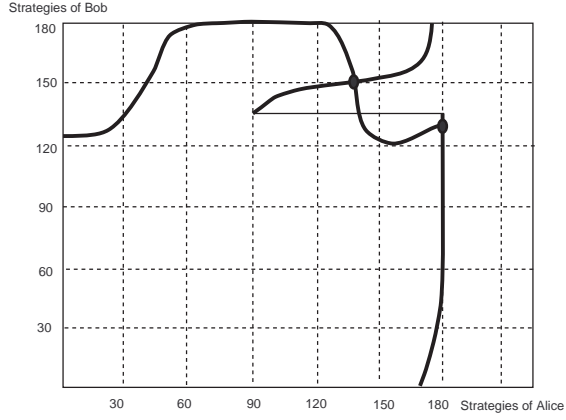}
\end{center}
\vspace{-15pt} \caption{\em Two points of Nash equilibrium}
\label{fig5}
\end{figure}
One of the equilibrium takes place for
$\alpha=145,5^0$,$\beta=149,5^0$ and gives the following
values for the squares of moduli of amplitudes:
\begin{center}
for Alice $p_1=0,679;\, p_2=0,509;\, p_3=0,321;\,
p_4=0,491;$

for Bob  $q_1=0,258;\, q_2=0,967;\, q_3=0,742;\,
q_4=0,033.$
\end{center}
\noindent The price of the quantum game, i.e. the
equilibrium value of the profit for Alice in this case is
equal to $~E\widehat{\cal H}_A=2.452~$. The second
equilibrium point corresponds to angles $\alpha=180^0$,
$\beta=123,5^0$ and the squares of the amplitude moduli
\begin{center}
for Alice $p_1=1.000; p_2=0,967; p_3=0.000; p_4=0.033;$

for Bob $q_1=0,695;q_2=0.646;q_3=0.305;q_4=0.354.$
\end{center}
\noindent The price of the game in the second equilibrium
point is equal to $~E\widehat{\cal H}_A=1.926~$.

{\bf 2. \em A unique equilibrium} is observed for example
in the case when all nonzero payoffs are equal and are
equal to one and for equal angles $\theta_A=45^0$,
$\theta_B=45^0$. The equilibrium point is located in the
upper right vertex of the square (see Fig.~\ref{fig6}):
\begin{figure}[ht]
\begin{center}
\includegraphics[height=0.48\textwidth]{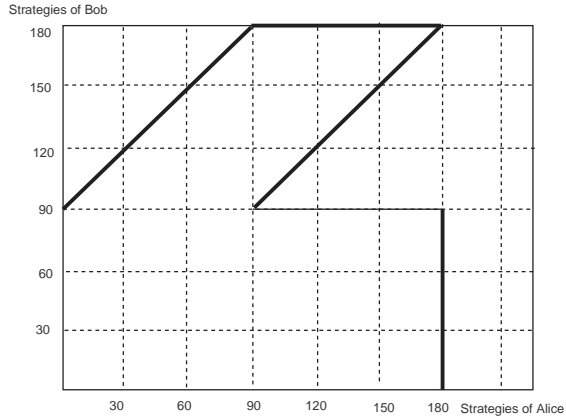}
\end{center}
\vspace{-15pt} \caption{\em The unique Nash equilibrium}
\label{fig6}
\end{figure}
The curve of Bob's reaction is shown on the Fig.~\ref{fig6}
as {\it continuous} while the analogous curve of Alice is
discontinuous when Bob is using the strategy corresponding
to the angle $\beta=90^0$.  To make it more explicit the
discontinuity is shown by drawing the thin line. In reality
{\it both} lines are discontinuous. This becomes evident if
one prolongs both functions on the whole real axis taking
into account the periodicity: the plots of one of them is
obtained by the shift of the other one on the halfperiod --
$90^0$. The squares of the amplitude moduli in this case
have the following values
\begin{center}
for Alice: $~p_1=1~$; $~p_2=0.5~$; $~p_3=0~$; $~p_4=0.5~$;

for Bob: $~q_1=1~$; $~q_2=0.5~$; $~q_3=0~$; $~q_4=0.5~$.
\end{center}
\noindent The payoff of the "wise" Alice in this case is
$~E\widehat {\cal H}_A=0.5~$.The unique equilibrium located
{\it inside} the square takes place for the initial payoff
matrix $a~=~3$,\; $b~=~3$,\; $c~=~5$,\; $d~=~1$ and angles
$\theta_A=15^0$, $\theta_B=35^0$ (see Fig.~\ref{fig7}).

\begin{figure}[ht]
\begin{center}
\includegraphics[height=0.48\textwidth]{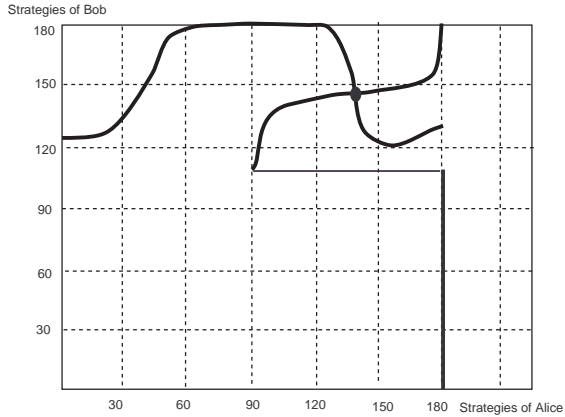}
\end{center}
\vspace{-15pt} \caption{\em The other example of the unique
Nash equilibrium} \label{fig7}
\end{figure}

\noindent {\bf 3. \em Absence of equilibrium}  is perhaps
one of the most interesting phenomena, because as it is
known for classical matrix games, equilibrium in mixed
strategies always exist. One can obtain absence of
equilibrium by taking the same payoff matrix for which one
as well as two points of equilibrium were found. For this
it is sufficient to take the operator representation of the
ortholattice with typical angles: $\theta_A=30^0$,
$\theta_B=20^0$.
\begin{figure}[ht]
\begin{center}
\includegraphics[height=0.48\textwidth]{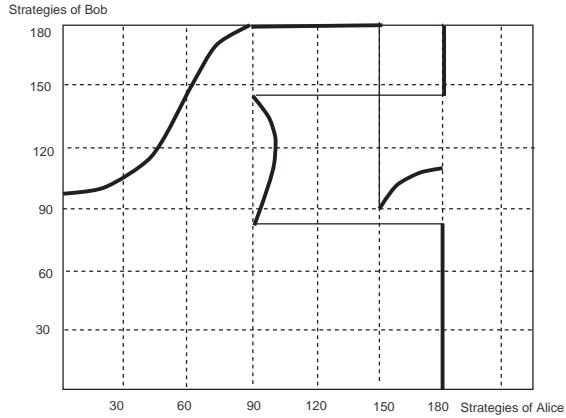}
\end{center}
\vspace{-15pt} \caption{\em Absence of Nash equilibrium}
\label{fig8}
\end{figure}
Absence of equilibrium in this case as it is seen from the
Fig.~\ref{fig8} is due to the discontinuity of the
functions of reaction which is impossible in the classical
case.We met this phenomenon in the first example when two
equilibrium points were obtained. This last example shows
the importance of the {\it realization} of a
nondistributive lattice. In the language of the game theory
one can understand it as follows: having the same interests
the players can form their behaviour qualitatively in
different ways. So the mathematician can give to the
client, for example to  Alice, strategic recommendations:
how she can organize the style of her behaviour to make the
profit larger for the same payoff conditions. For this,
however, he must know the choice of the representation of
Bob's logic.

\section*{Acknowledgements}
One of the authors (A.A.G.) is indebted to the Foundation
of the Ministry of Education of Russia, grant E0-00-14 for
the financial support of this work.

\end{document}